\documentstyle{mn}
\input psfig.tex

\newif\ifAMStwofonts

\ifoldfss
  \ifCUPmtlplainloaded \else
    \NewTextAlphabet{textbfit} {cmbxti10} {}
    \NewTextAlphabet{textbfss} {cmssbx10} {}
    \NewMathAlphabet{mathbfit} {cmbxti10} {} 
    \NewMathAlphabet{mathbfss} {cmssbx10} {} 
  \fi
  \ifAMStwofonts
    \ifCUPmtlplainloaded \else
      \NewSymbolFont{upmath} {eurm10}
      \NewSymbolFont{AMSa} {msam10}
      \NewMathSymbol{\upi}     {0}{upmath}{19}
      \NewMathSymbol{\umu}     {0}{upmath}{16}
      \NewMathSymbol{\upartial}{0}{upmath}{40}
      \NewMathSymbol{\leqslant}{3}{AMSa}{36}
      \NewMathSymbol{\geqslant}{3}{AMSa}{3E}

    \fi
  \fi
\fi 

\ifnfssone
  \newmathalphabet{\mathit}
  \addtoversion{normal}{\mathit}{cmr}{m}{it}
  \addtoversion{bold}{\mathit}{cmr}{bx}{it}
  \newmathalphabet{\mathbfit} 
  \addtoversion{normal}{\mathbfit}{cmr}{bx}{it}
  \addtoversion{bold}{\mathbfit}{cmr}{bx}{it}
  \newmathalphabet{\mathbfss} 
  \addtoversion{normal}{\mathbfss}{cmss}{bx}{n}
  \addtoversion{bold}{\mathbfss}{cmss}{bx}{n}
  \ifAMStwofonts
    \ifCUPmtlplainloaded \else
      \UseAMStwoboldmath
      \makeatletter
      \new@mathgroup\upmath@group
      \define@mathgroup\mv@normal\upmath@group{eur}{m}{n}
      \define@mathgroup\mv@bold\upmath@group{eur}{b}{n}
      \edef\UPM{\hexnumber\upmath@group}
      \new@mathgroup\amsa@group
      \define@mathgroup\mv@normal\amsa@group{msa}{m}{n}
      \define@mathgroup\mv@bold\amsa@group{msa}{m}{n}
      \edef\AMSa{\hexnumber\amsa@group}
      \makeatother
      \mathchardef\upi="0\UPM19
      \mathchardef\umu="0\UPM16
      \mathchardef\upartial="0\UPM40
      \mathchardef\leqslant="3\AMSa36
      \mathchardef\geqslant="3\AMSa3E
    \fi
  \fi
\fi 

\ifnfsstwo
  \DeclareMathAlphabet{\mathbfit}{OT1}{cmr}{bx}{it}
  \SetMathAlphabet\mathbfit{bold}{OT1}{cmr}{bx}{it}
  \DeclareMathAlphabet{\mathbfss}{OT1}{cmss}{bx}{n}
  \SetMathAlphabet\mathbfss{bold}{OT1}{cmss}{bx}{n}
  \ifAMStwofonts
    \ifCUPmtlplainloaded \else
      \DeclareSymbolFont{UPM}{U}{eur}{m}{n}
      \SetSymbolFont{UPM}{bold}{U}{eur}{b}{n}
      \DeclareSymbolFont{AMSa}{U}{msa}{m}{n}
      \DeclareMathSymbol{\upi}{0}{UPM}{"19}
      \DeclareMathSymbol{\umu}{0}{UPM}{"16}
      \DeclareMathSymbol{\upartial}{0}{UPM}{"40}
      \DeclareMathSymbol{\leqslant}{3}{AMSa}{"36}
      \DeclareMathSymbol{\geqslant}{3}{AMSa}{"3E}
    \fi
  \fi
\fi 

\ifCUPmtlplainloaded \else
  \ifAMStwofonts \else 
    \def\upi{\pi}
    \def\umu{\mu}
    \def\upartial{\partial}
  \fi
\fi

\def\simlt{\lower.5ex\hbox{$\; \buildrel < \over \sim \;$}}
\def\simgt{\lower.5ex\hbox{$\; \buildrel > \over \sim \;$}}

\def\zs{Z$_{\odot}$}

\begin{document}

\title{Metallicity in damped Lyman-$\alpha$ systems: 
     evolution or bias?}

\author[N. Prantzos and S. Boissier]
       {N. Prantzos and S. Boissier \\
 Institut d'Astrophysique de Paris, 98bis, Bd. Arago, 75104 Paris}
\date{ }

\pagerange{\pageref{firstpage}--\pageref{lastpage}}
\pubyear{1999}
\maketitle

\label{firstpage}

\begin{abstract}

Assuming that damped Lyman-$\alpha$ (DLA) systems are galactic discs, we
calculate the corresponding evolution of metal abundances. We use detailed
multi-zone models of galactic chemical evolution
(reproducing successfully the observed properties of disc galaxies) 
and appropriate statistics (including geometrical propability factors)
to calculate the average metallicity as a function of redshift. The results
are compatible with available observations, {\it provided that observational biases
are taken into account}, as suggested by Boiss\'e et al. (1998). 
In particular, high column density
{\it and } high metallicity systems are not  detected because the light 
of backround quasars is severely extinguished, while low column density
{\it and } low metallicity systems are not detectable through their 
absorption lines by current surveys. We show that these observational 
constraints lead to a ``no-evolution'' picture for the DLA metallicity, 
which does not allow to draw strong conclusions about the nature of those 
systems or about their role in ``cosmic chemical evolution''.

\end{abstract}

\begin{keywords}
Galaxies: general - evolution - spirals - photometry - abundances 
\end{keywords}

\section{Introduction}

Damped Lyman-$\alpha$ systems (DLAs) are high column density (N(HI)$>$10$^{20}$
cm$^{-2}$) absorbers detected in the optical spectra of quasars up to 
relatively high redshifts (up to $z\sim$5). Their study constitutes a powerful means
to investigate the properties of distant galaxies (or of their building blocks).
In  particular, DLA metal abundances have been widely used in the past few years
(e.g. Prochaska and Wolfe 1999, Pettini et al. 1999, Edmunds and Philips 1997,
 and references therein)
in order to probe the nature of DLAs (i.e. galactic discs vs. galaxies of other 
morphological types) or even to probe  the so-called ``cosmic chemical evolution''
(i.e. the global evolution of gas, metallicity and star formation rate 
in the Universe). It is not clear, however, whether the observed abundances
alone allow to probe the nature or the history of those systems. One reason is
possible depletion of metals into dust (e.g. Pei and Fall 1995). In this work
we investigate another factor, namely observational biases, along the
lines suggested by Boiss\'e et al. (1998). Assuming 
that (proto)galactic discs constitute a quite plausible model for such systems
we show that observational biases
may completely alter our interpretation of DLA metal abundances.

\section{Observational biases of metal abundance estimates in DLAs}

In Fig. 1 we present the empirical evidence on which this work is based.
Observations of Zn abundances in DLAs obtained by several groups are plotted
as a function of HI column density N(HI). This plot is an up-dated version of
Fig. 19 of Boiss\'e et al. (1998) and confirms the suggestion of those authors,
namely that there seems to be an anti-correlation between the observed Zn
abundance and N(HI), independently of redshift $z$.  As Boiss\'e at al. (1998)
notice, this is not to be interpreted as a physical correlation (i.e that
high metallicities are characteristic of low column density systems); indeed,
the observed abundance gradients in spiral galaxies offer clear evidence that
higher abundances are found in the inner disc, where gas column densities are
also higher (e.g. Garnett et al. 1997).

The correlation of Fig. 1 is rather to be interpreted in terms of observational
biases: no systems with a combination of metallicity and column density
outside the shaded region of Fig. 1 are presently detected, {\it even if such
systems do exist}.
The lack of high metallicity {\it and } high column density systems should be
attributed to extinction effects, since extinction depends on both those 
factors. The lack of low metallicity {\it and} low column density 
systems is  attributed by Boiss\'e et al. (1998) again to observational 
selection effects, i.e. to  the fact that below some level the amount of Zn
atoms along the line  of sight is insufficient to allow for a proper 
spectroscopic detection.

How ``realistic'' are these empirically determined detectability constraints
and how can they be quantitatively understood?
We postpone a  discussion of those matters to Sec. 4.
We simply wish to illustrate here the effect of these constraints
(assuming they are real)
on the detectability of a Milky Way type disc 
through its metal absorption lines.

In a recent work (Boissier and
Prantzos 1999, herefter BP99) we presented a detailed model for the chemical and 
spectrophotometric evolution of the disc of our Galaxy.
The galactic disc is simulated as an ensemble of concentric, independently
evolving rings, gradually built up by infall of primordial composition. The
chemical evolution of each zone is followed by solving the appropriate
set of integro-differential equations. 
The spectrophotometric evolution is followed in a self-consistent way, i.e.
with the  star formation rate  $\Psi(t)$ and metallicity $
Z(t)$ of each zone determined 
by the chemical evolution, and the same stellar Initial Mass Function
(from Kroupa et al. 1993). 
The adopted stellar yields, lifetimes, evolutionary tracks and spectra are
metallicity dependent.
 Dust absorption is included according to the prescriptions of  
Guiderdoni et al. (1998) and assuming a ``sandwich''
configuration for the stars and dust layers (Calzetti et al. 1994).
The star formation rate (SFR) is locally given by a
Schmidt-type law, i.e proportional to some power of the gas surface
density $\Sigma_g$ and varies with galactocentric radius $R$ as:
\begin{equation}
 \Psi(t,R) \ = \  \alpha \  \Sigma_g(t,R)^{1.5} \ V(R) \ R^{-1}
\end{equation}
where $V(R)$ is the circular velocity at radius $R$. This radial dependence of
the SFR is suggested by
the theory of star formation induced by density waves in spiral
galaxies (e.g. Wyse and Silk 1989).

It turns out that
the number of observables reproduced by the model is much
larger than the number of free parameters. In particular		
the model reproduces present day ``global'' properties 
(amount of gas, SFR, and supernova rates), as well as	
the current Milky Way disc luminosities in various wavelength bands 
and the corresponding radial profiles of gas, stars, SFR and metal abundances;
moreover, the adopted inside-out star forming scheme leads to a 
scalelength of $\sim$4 kpc in the B-band and $\sim$2.8 kpc in the K-band, 
in agreement with observations (see BP99).

At this point it should be noticed that, among all metals observed in DLAs
through absorption line measurements,
Zn is usually considered to be the most reliable tracer of metallicity,
because its abundance  is expected to suffer little from depletion
into dust (e.g. Pettini et al. 1997)
However, from the theoretical point of view, the nucleosynthesis
of Zn is not well understood. The only known production site
is massive stars, and the most detailed  models of this site are 
those of Woosley and Weaver (1995, WW95),  who give the only
available models with yields of the various isotopes as a function of
initial stellar metallicity. The Zn yields of these models show an
unexplained behaviour: the dominant Zn isotope, $^{64}$Zn,
is underproduced, while the yields of the next two more important isotopes,
$^{66}$Zn and $^{68}$Zn, increase strongly as  metallicity increases from
0.1 \zs \ to \zs. The reason for this behaviour in the models is not  yet well
understood. As a result, the [Zn/Fe] ratio calculated with the
metallicity dependent yields of WW95 and a chemical evolution model
shows a rather abrupt and pronounced increase for metallicities [Fe/H]$>-1,$  
while it stays nearly constant
at lower metallicities (Timmes {\it et al.} 1995a, 
Goswami and Prantzos 2000). This behaviour is not found
in the observed pattern of the Zn/Fe ratio: both halo and disk stars of all
metallicities show solar Zn/Fe (see Goswami and Prantzos 2000 and references
therein), despite the fact that SNIa, the main Fe producers in the disk,
do not produce significant Zn amounts.

\begin{figure}
\psfig{file=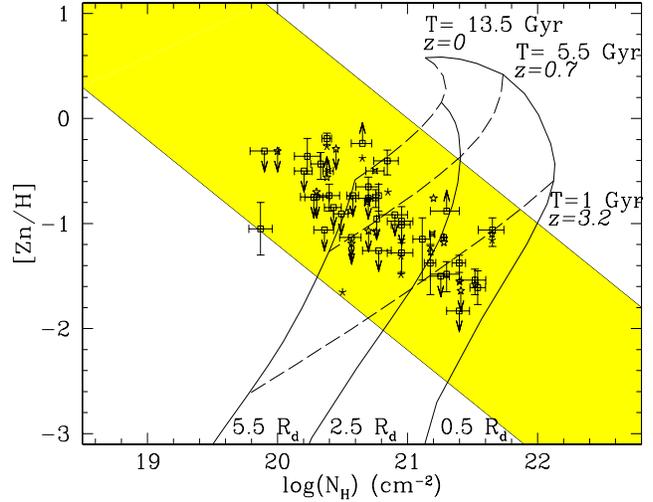,height=7.cm,width=0.5\textwidth,angle=-90}
\caption{\label {}
Zn abundances vs gas column density N(HI). 
Observations in DLAs are from Pettini et al. 
(1994, 1997, 1999), 
Prochaska and Wolfe (1999), 
Boiss\'e et al. (1998),
Lu et al. (1996).
Observations are apparently contained within the {\it shaded area}, 
which is limited by
[Zn/H]+log(N(HI))$<$21 ({\it upper diagonal}) and  
[Zn/H]+log(N(HI))$>$18.8 ({\it lower diagonal}).
For illustration purposes, we show
the evolution of Zn vs gas column density in three different
zones of a Milky Way type disc ({\it solid curves}, 
at 0.5, 2.5 and 5.5 scalelengths
from the center, respectively) according to our models. 
The column densities correspond to our model disc seen ``face-on''.
The disc
evolves from the lower left to the upper right in this diagram, as can be seen
from the 3 isochrones ({\it dashed curves}) 
at times T=1 Gyr, 5.5 Gyr and 13.5 Gyr, respectively; the corresponding redshifts
are for a cosmological model with $H_0$=65 km/s/Mpc, $\Omega_0$=0.3 and a galaxy 
formation redshift $z$=6 (i.e. the time of formation of the first stars). 
}
\end{figure}

This situation does not allow us to trust theoretical nucleosynthesis
prescriptions for the evolution of Zn, since theory seems unable
to reproduce  observations in the Galaxy.
Therefore we adopt an empirical approach for the nucleosynthesis of Zn,
already suggested in Timmes {\it et al.} (1995b): based on the observed
Zn/Fe pattern in the Galaxy, we assume that Zn traces Fe at all metallicities.
Since the history of Fe in the Galaxy
is observationally well constrained (at least in the solar neighborhood)
and reasonably well reproduced by our models (see BP99), 
we assume that it can be safely used to
trace the history of Zn in other places as well. 

In Fig. 1 we present the evolution of the Zn abundance in three representative
zones of the Milky Way model (corresponding to the inner disc, the solar
neighborhood and the outer disc, respectively) as a function of the
corresponding gas column densities of those zones. It should be noticed that
in our models we do not distinguish between atomic and molecular gas and the 
curves in Fig. 1 are drawn by assuming that all the gas is in atomic form
(the contribution of He, i.e. 10\% by number, is properly removed). This
approximation certainly affects the results (i.e. the curve corresponding to 
the inner disc, where most of the gas is known to be 
in molecular form, should be shifted 
to the left by 0.5 on that same scale), but for our 
illustration purposes the figure is quite appropriate.

Indeed, it is clearly seen that the actual chemical evolution of the disc cannot be revealed
by observations, because of the empirically determined constraints discussed previously.
At any given epoch (i.e. along the ``isochrones'' in Fig. 1) only a sub-sample of the
disc can be probed. At early times this sub-sample is representative of  the inner regions
which reach rapidely relatively high column densities and moderate metallicities; 
at late times these regions develop such
high densities {\it and} metallicities that they move into the 
``forbidden'' part of the diagram (upper right).
At late times then, it is the outer regions that 
can be probed through the absorption lines.

Obviously, in such conditions it is difficult to derive the real chemical history of the 
system, since only snapshots of {\it different regions at different times} can be available,
not of the same regions at different times.

\section{Metallicity evolution in DLAs}

The discussion in the previous section suggests that our picture of DLA chemical evolution,
obtained through quasar absorption line measurements, may be seriously biased. 
Indeed, if taken at face value, Fig. 1 suggests that at early times the lowest metallicities and
at late times the highest metallicities are unobservable. Could this
bias modify the picture to such an extent as to give the impression of no-evolution at all?

\begin{figure*}
\psfig{file=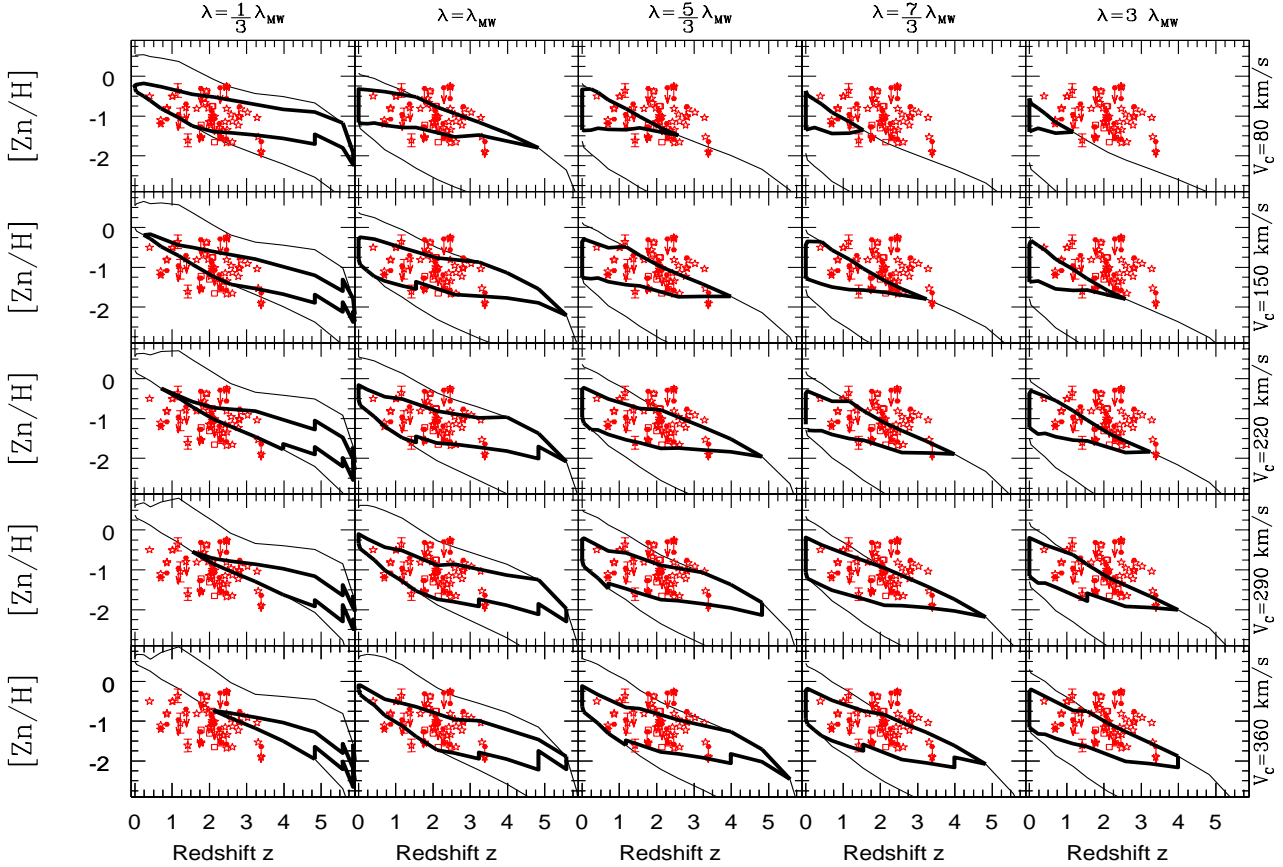,height=12.cm,width=\textwidth,angle=-90}
\caption{\label {}
Evolution of Zn abundances in our models as a function of redshift $z$. Results are
presented for a grid of 25 disc models, caracterised by 5 values of
the disc maximal circular velocity V$_C$ (80, 150, 220, 290, 360 km/s, {\it from top to bottom})
and 5 values of the spin parameter $\lambda$/$\lambda_{MW}$ (1/3, 1, 5/3, 7/3, 3,
{\it from left to right})
where $\lambda_{MW}$ is the corresponding value for the Milky Way disc.
For clarity, only the evolution of two ``extreme'' zones is shown for each model, at 0.5
and 5.5 scalelengths from the center ({\it upper and lower  curve}, respectively,
in each panel), which span the evolution of the whole disc. 
The region enclosed within {\it thick curves} is obtained by application of the
empirical selection criterion of Fig. 1 (i.e.
18.8 $<$ [Zn/H]+log(N(HI)) $<$ 21), assuming that all discs are seen ``face-on''.
It can be seen that this ``filter'' selects
zones within a restricted range of [Zn/H] value (independently of redshift), and leads
to a no-evolution picture. Observations (the same in all panels) 
are from the references listed in the legend of Fig. 1.}

\end{figure*}

In order to answer quantitatively this question, a working model for the evolution of DLAs
is needed. Some authors suggested that DLAs are proto-galactic discs (e.g. Prochaska and Wolfe
1996, Ferrini et al. 1997, Prantzos and Silk 1998) 
while others interpreted them as low surface brightness galaxies (Jimenez
et al., 1998), dwarf irregulars (Matteucci et al. 1997), galactic halos (Valageas et al. 1999)
 or proto-galactic
``building blocks'' (Haehnelt et al. 1998, Ledoux et al. 1998). 
It has also been suggested that DLAs differ substantially from the galaxies that
contribute mostly to the observed SFR in the Universe (e.g. Pettini et al. 1999).

We shall adopt in this work the hypothesis that DLAs are galactic discs.
We shall use a detailed model we developed recently for the chemical and
spectrophotometric evolution of galactic discs
(Boissier and Prantzos 2000). It is essentially the
Milky Way model presented in Sec. 2, extended to disc galaxies through
``scaling properties'' derived by Mo, Mao and White (1998)
in the framework of the Cold Dark Matter (CDM) scenario for galaxy formation. 
In our simplified version of this scenario
disc profiles can be expressed in terms of only two 
parameters: maximal rotational velocity $V_C$ 
(measuring the mass of the halo and, by assuming a 
constant halo/disc mass ratio, also the mass of the disc) and spin 
parameter $\lambda$ (measuring the specific angular momentum of the halo).
If all discs are assumed to start forming their stars at the
same time (but not at the same rate!),
the profile of a given disc can  be expressed in terms of the one of our
Galaxy.
We  constructed a grid of 25 models caracterised by $V_C$ = 80, 150, 220, 290, 360 km/s
and $\lambda/\lambda_{MW}$ = 1/3, 1, 5/3, 7/3, 3,  respectively,  where $\lambda_{MW}$
is the spin parametre of the Milky Way.
Increasing values of $V_C$ correspond to more massive discs and
increasing  values of $\lambda$ to more extended ones. 
The SFR is calculated from Eq. 1, with the appropriate velocity profile $V(R)$.
Notice that
the efficiency $\alpha$ is not a free parameter, since it is the same as in
the Milky Way model.

It turns out that this simple model  reproduces fairly well most of the main
properties of present days discs (Boissier and Prantzos 2000):
disc sizes and central surface brigthness, Tully-Fisher relations in various 
wavelength bands, colour-colour and colour-magnitude relations, gas fractions 
vs. magnitudes and colours, abundances vs. local and integrated properties,
as well as integrated spectra for different galactic rotational velocities. 
Moreover, as shown in Prantzos and Boissier (2000), it also reproduces the 
observed abundance gradients in disc galaxies.

In Fig. 2 we present the results of our models for the Zn evolution in DLAs
(assumed to be galactic discs) as a function of redshift $z$. For clarity, only the
evolution of the inner and outer disc is presented in each panel 
({\it thin curves}, corresponding
to zones located at 0.5 and 5.5 scalelengths from the centre, respectively).
Star formation is assumed to start at redshift $z$=6 for all discs, but any value
of $z>$4-5 would produce results similar to those displayed here.
The resulting evolution is not very different from that calculated in e.g. Prantzos
and Silk (1998) for the Milky Way, or in Ferrini et al. (1997) with 
multi-zone disc models.
Notice that Malaney and Chaboyer (1996), Timmes et al. (1995b), Matteucci et al.
(1997), Edmunds and Philipps (1997) and Lindner et al. (1999) 
have studied metallicity evolution in DLAs with
one-zone models, differing by the
star formation timescales or by the time of the beginning of star formation.

It can be clearly seen in Fig. 2 that between redshifts $z\sim$3 and $z\sim$1
there is substantial metallicity evolution in all galactic zones, typically an increase
by a factor $\sim$10.
Such an increase is certainly not observed in the available data, which is also displayed
on each one of the panels in Fig. 2.

The region enclosed within {\it thick solid curves} in Fig. 2 is obtained by application
of the ``empirical constraints'' of Fig. 1, i.e. 
by excluding all regions with a combination of
metallicity and column density $F$(Zn,N(HI)) = [Zn/H] + log(N(HI)) such that $F<$18.8 or
$F>$21. The resulting  observational picture is now completely different from the real one:
no sizeable evolution in metallicity is observed (except, perhaps, in the lowest
redshift  range, where metallicities are somewhat higher than average).

It should be noticed that the empirical ``filter'' has been applied to our models
by assuming that discs are seen ``face-on'' and that all the gas is in the form of
HI. These simplifying assumptions have opposite effects on the derived
column density along the line of sight: adopting
a different inclination would increase the column density of our disc models; 
taking into account that part of the gas is in the form of H$_2$, would decrease it.
The former factor can be treated statistically, but not the later. Taking all other 
uncertainties into account (i.e. possible variations in the 
H$_2$/HI ratio with metallicity, see
Combes 1999) we think that Fig. 2 gives a rather good first approximation to the real
situation.

The results of Fig.2 are summarised in the upper panel of Fig. 3, wher we plot in the
same diagram all the ``filtered'' zones of our models as a function of redshift
({\it shaded area}) and compare them to observations. The ``no-evolution'' picture
is even more clearly seen now, especially in the $z\sim$1-3 redshift range.
A firm prediction of these models is that the Zn abundances of DLAs at higher
redshifts, in the range $z\sim$3-5, will be not too different from  those  already detected
at $z\sim$1-3.

Assuming that this first approximation is correct (i.e. that DLAS are indeed galactic discs)
it is interesting to calculate the {\it most probable metallicity values} expected at
a given redshift. Even if our basic assumptions are correct, this is by no means 
a trivial task, since it implies the knowledge of the appropriate statistical factors
as a function of the redshift. For illustration purposes we adopt here the following
simplified assumptions:

i) the distribution function of discs in the velocity space $V_C$ is time-independent and
given by the expression suggested in Gonzalez et al. 1999 (in the following we simplify,
for clarity, the notation $V_C$ to $V$, unless if explicitly stated otherwise):
\begin{equation}
F_V(V) dV = \tilde{\Psi}_* \left(\frac{V}{V_*} 
\right)^{\beta}exp\left[-\left(\frac{V}{V_*}\right)^n\right]\frac{dV}{V_*}.
\end{equation}
\noindent
The parameters $\Psi_*, \ V_*, \ \beta$ and $n$ are  determined  in Gonzalez et al. (1999)
on the basis of observed Tully-Fisher relationships
and luminosity (Schechter-type) functions. 
We adopt here the set of parameters of their Table 4 (fifth row, LCRS-Courteau data) 
corresponding to the velocity interval covered by our models.
Our results would not be affected much by the choice of another velocity function, since
the form of $F_V$ always favours discs of low $V$.

ii) the distribution function in spin parameter $\lambda$-space is 
time-independent and given by:
\begin{equation}
F_{\lambda}(\lambda)d\lambda \ = \ \frac{1}{\sqrt{2\pi}\sigma_{\lambda}} \
exp \left[ - \frac{ln^2(\lambda/\bar{\lambda})}{2 \sigma_{\lambda}^2} 
\right]\frac{d\lambda}{\lambda}
\end{equation}
with $\bar{\lambda}$=0.05 and $\sigma_{\lambda}$=0.5 (obtained by numerical 
simulations, see e.g. Mo, Mao and White, 1998) and $\lambda_{MW}$=0.06 for the
Milky Way disc (Sommer-Larsen, private communication).
The $\lambda$-function favours moderately ``compact'' discs
(those with $\lambda\sim$0.04-0.05).

iii) the probability that a line of sight to a QSO intercepts a disc
in the radius interval $[R,R+dR]$ is proportional to the geometrical
cross-section $F_R dR =  2 \pi R dR$,
favouring the detection of the outer regions of the larger discs. 

iv) the distributions $F_V$, $F_{\lambda}$ and $F_R$ are independent.

Applying the joint probability function $F(V,\lambda,R) = F_V F_{\lambda} F_R$
to our models, we obtain the {\it mean metallicities} $<[Zn/H]>$ shown in Fig. 3.
The two curves in the upper panel are obtained by:
\begin{equation}
\left< \left[ 
\frac{Zn}{H} \right] \right> = \frac{\int_{\lambda}\int_{V} \int_0^{R_{L}} 
F(\lambda,V,R) \Phi(R) [\frac{Zn}{H}(R)] \ dR \ dV \ d\lambda}
{\int_{\lambda}\int_{V}\int_0^{R_{L}} F(\lambda,V,R) \Phi(R) \ dR \ dV \ d\lambda }
\end{equation}
\noindent
where $R_{L}$ is the radius of the largest disc in our models. The mean value over the
whole disc (i.e. without applying the empirical ``filter'') corresponds to $\Phi(R)$=1 in
all zones and is given by the thin curve in Fig. 3. In that case, $<[Zn/H]>_U$ 
($U$ for ``Unfiltered'') increases by a factor
$\sim$20 between redshifts $z$=3 and $z$=1 and  is clearly below all observational data  (because
the outer, low metallicity, regions of the discs are favoured in that case). This shows the
importance of properly taking into account various statistical factors, something that has not
been done in previous studies of DLAs with multi-zone disc models
(e.g. in Ferrini et al. 1997 and Prantzos and Silk 1998).

\begin{figure}
\psfig{file=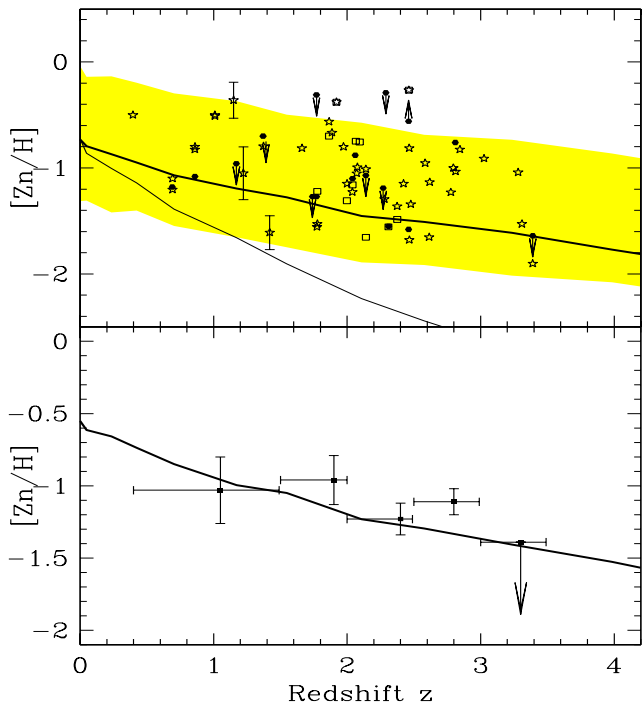,height=10.cm,width=0.5\textwidth}
\caption{\label {}
Evolution of {\it mean metallicities} of our disc galaxy models as a function of
reshift $z$. {\it Upper panel:} The {\it thin curve} is the mean metallicity of all
the zones of our models, i.e. the ``unfiltered'' value $<[Zn/H]>_U$ obtained with
$\Phi(R)$=1 always in Eq. (4); the {\it thick curve} is the mean value $<[Zn/H]>_F$
of the zones ``filtered'' through the empirical constraint of Fig. 1, i.e. the mean value
of the shaded area, obtained with $\Phi(R)$=1 for those filtered zones and $\Phi(R)$=0
for the others. The {\it shaded area} includes all the ``filtered'' zones of our models,
i.e. all the zones inside {\it thick curves} in Fig. 2. 
Observations are as in Fig. 2. {\it Lower panel}: Column density
weighted average metallicity of the ``filtered zones'' $<[Zn/H]>_{FW}$, obtained by
Eq. (5) ({\it thick curve}); data are from Pettini et al. (1999).   }
\end{figure}

The mean value $<[Zn/H]>_F$ over the ``filtered' disc zones (i.e. those in the shaded aerea
of Fig. 3) is obtained with $\Phi(R)$=1 in those zones and $\Phi(R)$=0 outside them.
It is shown by  the {\it thick curve} in Fig. 3.
This  $<[Zn/H]>_F$ value is in the lower range of the ``filtered'' values,
again because of the geometrical factor $F_R$.
$<[Zn/H]>_F$ increases by a factor of $\sim$2 between 
$z$=3 and $z$=1, an increase which is compatible with the observations.

Pettini et al. (1999) have also estimated the column density weighted average of the
[Zn/H] values in DLAs, by binning their data in 5 redshift bins (the last bin, at $z>3$,
being in fact an upper limit only). Again, no substantial evolution is seen
in the data ({\it lower panel } in Fig. 3). We  also
calculated the corresponding average metallicity in our models, by folding with
the gas column densities $N_H(R)$ of the ``filtered'' zones (assuming that  all 
but 10 \%  by number - corresponding to He - is in the form of atomic hydrogen):
\begin{equation}
\left< \left[ 
\frac{Zn}{H} \right] \right>_{FW} = \frac{\int_{\lambda}\int_{V} \int_0^{R_{L}} 
F(\lambda,V,R) \Phi \ [\frac{Zn}{H}]  \ N_H \ dR \ dV \ d\lambda}
{\int_{\lambda}\int_{V}\int_0^{R_{L}} F(\lambda,V,R) \Phi \ N_H \ dR \ dV \ d\lambda }
\end{equation}
\noindent
where $\Phi$, [Zn/H] and $N_H$ depend on radius $R$.
The resulting evolution is shown in the lower panel of Fig. 3. It can be seen that the
weighted mean metallicity  $<[Zn/H]>_{FW}$
 of the ``filtered'' zones of our models evolves in a way which is certainly compatible with
the data. A clear prediction of the model is that at low redshifts ($z<$1) there should be as
much evolution as in the $z$=1-3 range (i.e. a factor of $\sim$2-3 increase in the
weighted mean metallicity of DLAs in both cases).

In summary, by using a self-consistent model of galactic chemical evolution
(i.e. reproducing in detail the properties of local galaxies) and incorporating
``reasonable'' statistics and appropriate empirical constraints, we have 
shown unambiguously that only a small degree of evolution should be expected for the
obeserved mean metallicity of DLAs; according to our models, these systems may well be
galactic discs.

\section {Origin of biases}

It is interesting to see  how the empirically determined upper limit in the Zn vs N(HI)
plane may be intrepreted in terms of extinction. 
We shall assume here that our model discs
constitute gaseous screens, seen face-on and reducing  the intensity of the light of
background quasars. The corresponding extinction is calculated 
in the rest-frame of the absorber by the formula:
\begin{equation}
A_{\lambda} (z) \ = \ (A_{\lambda}/A_V)_{\odot} \ 
(A_V/N_H)_{\odot} \ N_H \ (Z/Z_{\odot})^{1.6} 
\end{equation}
where $(A_{\lambda}/A_V)_{\odot}$ is the local normalised extinction curve 
(Natta and Panagia,1984),
$(A_V/N_H)_{\odot}$ is the extinction in the V-band 
in the solar neighborhood (from Bohlin et al., 1978)
and the exponent 1.6 in the metallicity term is introduced in order to reproduce the observed
extinction curves in the Small and Large Magellanic Clouds 
(Guiderdoni and Rocca-Volmerange, 1987).
The wavelength $\lambda$ depends on the redshift $z$ of the absorber. Since the
background QSO light is observed in the visible ($\lambda_{V,obs}$=0.55 $\mu$), we
have for the corresponding wavelength $\lambda_{ABS}$ of the absorber
\begin{equation}
\lambda_{ABS} (z) \ = \ \frac{\lambda_{V,obs}}{1+z} 
\end{equation}

\begin{figure}
\psfig{file=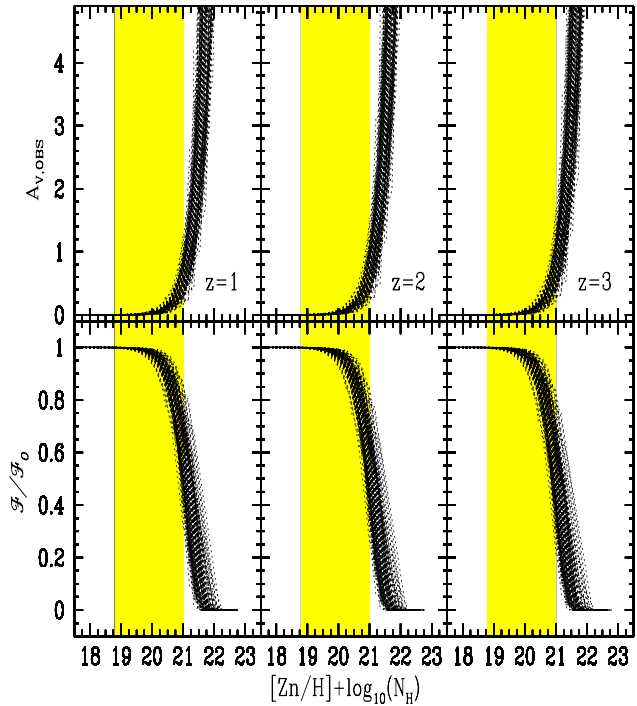,height=10.cm,width=0.5\textwidth,angle=-90}
\caption{\label {}
{\it Upper panel}:
Extinction A$_V$ of background light caused by the various zones of our models, 
assumed to be gaseous screens
(seen face-on) at redshift $z$=1,2,3 (from left to right, respectively) of 
column density N(H) and metallicity [Zn/H] (Eq. 6) 
{\it Lower panel}: Corresponding reduction factor in the brightness of the background
light. In both panels the {\it shaded area} corresponds to the one of Fig. 1, i.e.
satisfying the ``filter'' 18.8 $<$ [Zn/H]+log(N(H)) $<$ 21. It is clearly seen that
extinction increases rapidly to the right of the upper limit, presumably making
the bakground QSOs unobservable. 
}

\end{figure}

For illustration purposes, we calculated the extinction at 3 different redshifts
$z$=1,2,3 (i.e. in the absorber's rest-frame $\lambda_{ABS}$= 0.275, 0.183 and 0.137 $\mu$,
respectively) during the evolution of all the zones of our models. We
plot it in Fig. 4 as a function of $F$=[Zn/H]+log(N(HI)) ({\it upper panel}), while in the lower
panel we plot the corresponding fraction of background light filtered through the screen. 
It can be seen that for values of $F>$21 (i.e. above the empirically
determined limit in Fig. 1) extinction increases rapidly, reaching
1 mag at $z$=1, 1.5 mag at $z$=2 and 2 mag at $z$=3). The background intensity drops below
40\% of its initial value. Our results substantiate the claim of Boiss\'e et al. (1998) that
extinction is biasing the interpretation of metallicity abundance determinations in DLAs.

As for the lower value of the ``empirical'' constraint, it can be understood as follows:
Taking into account that, by definition, DLAs correspond to column densities
log(N(HI)) $>$20, the lower limit $F<$18.8 corresponds to [Zn/H]$<$-1, i.e. to
less than 5 10$^{11}$ atoms of Zn per cm$^2$
along the line of sight (adopting a solar
ratio (Zn/H)$_{\odot}\sim$4.5 10$^{-8}$ by number, e.g. Anders and Grevesse 1989). 
Current surveys of DLAs typically reach N(Zn)$>$10$^{12}$ cm$^{-2}$ (Pettini et al. 1997),
which explains, perhaps, why the lower left part of  Fig. 1 is void.

\begin{figure}
\psfig{file=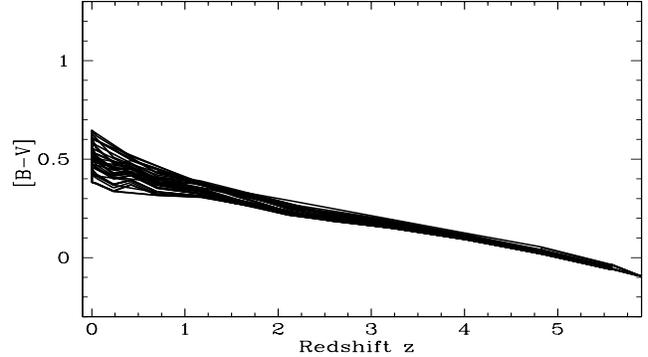,height=5.cm,width=0.5\textwidth,angle=-90}
\caption{\label {}
B-V vs. redshift for the stellar populations of the disc zones of our models
that satisfy the condition
18.8 $<$  [Zn/H] + log(N(HI)) $<$ 21 (corresponding to the shaded aerea of Fig. 3).
Discs are assumed to be seen ``face-on''; inclined discs should provide
higher B-V values.
}

\end{figure}

In any case, it should be interesting  to detect  the stellar 
population responsible for
the chemical enrichment of the gas in DLAs. 
In Fig. 5 we plot our model B-V values of that 
population, which span a narrow range 0.2$<$B-V$<$0.4 for redshifts 1$< z <$3.
In fact, these values are lower limits, since our model 
discs are assumed to be seen face-on.
Thus, one of the firm predictions of our model applied to DLAs is that the
underlying stellar population should be somewhat redder than the curve in Fig. 4.

\section {Summary}

The observationally determined constraint 18.8 $<$ [Zn/H]+log(N(HI)) $<$ 21 
(Boiss\'e et al. 1998)
has profound implications for the intrepretation of metal abundances detected in DLAs.
Assuming that DLAs are galactic discs, and using detailed (and successful) models
for disc evolution, we show that current observations cannot probe the true evolution
of those systems: a no-evolution picture, compatible with available data, emerges
when the empirical constraints are taken into account. 
We calculate average metallicities in the galactic zones of our models that are
``filetered'' by the empirical constraints, by taking into account appropriate
statistical factors. We find that the resulting 
column density weighted average metallicity shows a small increase at low redshifts and
is compatible with currently available data. These findings suggest that DLAs 
may well be galactic discs, as argued e.g. in Prochaska and Wolfe, 1999 
(for a different view, see Pettini et al., 1999).

We also show quantitatively how
extinction may be indeed responsible for the non-detection of metal rich DLAs of high column
density, as suggested by Boiss\'e et al. (1998); according to those authors, such DLAs
may be one day detected in samples drawn from the observation of fainter QSOs, independently
of the redshift. Moreover, metal poor DLAs of low column density should also be detected
with more sensitive instruments. Finally, if our interpretation of currently observed
DLAs as galactic discs
is correct, we expect that the underlying stellar populations should have
B-V$>$0.2 (by a small amount) in the redshift range 1$< z <$3.

\bigskip
\noindent
{\bf Acknowledgements:} We are grateful to Patrick Boiss\'e and Patrick Petitjean for useful discussions
and comments on this work.

\label{lastpage}

\end{document}